# Correlated disorder in myelinated axons orientational geometry and structure

**Michael Di Gioacchino** [1,2*], **Gaetano Campi** [1], **Nicola Poccia** [3], **Antonio Bianconi** [1,2]

[1] Institute of Crystallography, CNR, Via Salaria Km 29.300, Monterotondo, Roma, I-00015, Italy.;
gaetano.campi@ic.cnr.it

[2] Rome International Center of Materials Science (RICMASS), Via dei Sabelli 119A, 00185 Roma, Italy;
antonio.bianconi@ricmass.eu

[3] Department of Physics, Harvard University, Cambridge, Massachusetts 02138, USA; npoccia@g.harvard.edu

[*] Correspondence: michael.digioacchino@gmail.com; Tel.: +39-06-9067-2624

**Abstract:** While the ultrastructure of the myelin has been considered to be a quasi-crystalline stable system, nowadays its multiscale complex dynamics appears to play a key role for its functionality, degeneration and repair processes following neurological diseases and trauma. In this work, we have investigated the interactions between axons associated to the nerve functionality, measuring the spatial distribution of the orientation fluctuations of axons in a *Xenopus Laevis* sciatic nerve. At this aim, we have used scanning micro X-ray diffraction (SµXRD), a non-invasive technique already applied to other heterogeneous systems presenting complex geometries from microscale to nanoscale. We have found that the orientation spatial fluctuations of fresh axons show a Levy flight distribution which is a direct indication of correlated disorder. We have found that the Levy flight distribution is missing in the aged nerve prepared in the unfresh state. This result shows that the spatial distribution of axons orientation fluctuations in unfresh nerve state loose the correlated disorder assuming the random disorder behavior. This work allows a deeper understanding of the ultrastructure-function nerve relation and paves the way to study other materials and biomaterials using SµXRD technique to detect fluctuations of their supramolecular structure.

## 1. Introduction

The myelin is a multilamellar insulating structure that wraps axons of some vertebrate neurons both in the central and peripheral nervous system (CNS, PNS). It has as main functions the protective, nourishing and the enhancement of the transmission of the nerve impulses, through the saltatory conduction [1,2].

We focus our attention on the myelin of a sciatic nerve, member of PNS, of a frog *Xenopus Leavis*. In PNS, the myelin sheath is formed by the spiral winding of the Schwann cells (Sc), around the axon. It has a periodic multimellar structure whose unit is made of the succession of four layers [1]: cytoplasmatic apposition of Sc (*cyt*), lipidic membrane (*lpg*) extracellular apposition of Sc (*ext*) and another lipidic membrane (*lpg*). Many information about these layers has been extracted from measure of space averaged laboratory X-ray diffraction (XRD) [3-5] and electron microscopy (EM) [6,7] and neutron diffraction [8]. The study of myelin, using its diffractive nature, has a long history of experiments from the early 1950s to today. These measurements have allowed to determine the thickness of the layers under various experimental conditions however experimental methods probing the reciprocal k-space with high resolution, like x-ray or neutron diffraction, have no spatial resolution therefore have provided the myelin quasi-crystalline lattice structure information averaged on sections of the sample as large as the incident beam size.

The missing information on the spatial fluctuations of the myelin ultrastructure can be now investigated using the novel technique scanning micro X ray diffraction (SµXRD) [8-15] which allow to probe both the real space and the k-space with high resolution thanks to the ability to focus X ray beams on micron size on biological samples, paving the way to the new field of statistical spatial fluctuations of biological systems (tissues, fibres, etc.) which can be investigated from micro-scale down to nanoscale, beyond the known averaged structure seen at larger scale. Moreover, in case of myelin, there is great interest around the study of the dynamical structural fluctuations of myelin proteins [8], and composition [11,12]. In this context, this new approach can be the starting point for the treatment of damage and degeneration of the myelin sheath, like demyelization.

In this work we have focused our attention on the interactions between axons in the sciatic nerve, associated to the nerve functionality, measuring the spatial distribution of the orientation fluctuations of axons. About the axons orientation, there are many techniques that allow the visualization of isolated axons such as X-ray fluorescence microscopy, atomic force microscopy, scanning and transmission electron microscopy. However, up to now, no quantitative measurements of the fluctuations of axons orientation within the nerve have been reported. Here we have evaluated the axons orientation and its fluctuations into the nerve, using SµXRD on the sciatic nerve myelin of a frog *Xenopus Laevis*. SµXRD technique has already been used for studying biomaterials with high spatial resolution, such as bones, tissues and cells [9-15]. We measured axons myelin orientation point by point, using synchrotron radiation X-ray beam focused to micrometric dimensions at the ID 13 beamline at European synchrotron radiation facility, ESRF, in Grenoble [9-15]. Afterwards, we applied statistical physics tools to the big data set of diffraction patterns collected to unveil the "statistical distributions of the fluctuating angular order".

Our results indicate that *i*) the distribution of the orientation fluctuations of the axons in the fresh functional sciatic nerve shows a Levy probability distribution with large fluctuations typical of quantum matter and complex systems [16-21]; *ii*) the axons angular distribution becomes randomly distributed in the early stages of degeneration of myelin following an 18h ageing in a culture, at room temperature without oxygenation and insertion of ATP. This loss of structural correlated disorder is accompanied by an increased rigidity of the system. The results show that our statistical analysis of SµXRD appears to be a powerful tool to determine spatial fluctuations of different functional states of biological matter.

**2. Materials and Methods**

2.1 Samples preparation.

*Xenopus Laevis* adult frogs (12 cm length, 180-200 g weight, *Xenopus* express, France) were housed and euthanized at the Grenoble Institute of Neurosciees. The local committee of Grenoble Institute of Neurosciences approved the animal experimental protocol. The sciatic nerves were ligated with sterile silk sutures and extracted from both thighs of freshly sacrificed *Xenopus* frog at approximately the same proximal-distal level through a careful dissection of the thigh.

The fresh state of the sciatic nerves was obtained by placing the nerve after dissection immediately in a culture medium at pH 7.3 in a thin-walled quartz capillary, sealed with wax for the SµXRD imaging measurements.

The unfresh state of sciatic nerves was obtained by keeping the nerve for 18 hours after dissection at room temperature, without oxygenation and without insertion of ATP in Petri dish equilibrated in culture medium at pH 7.3, and it was prepared for the SµXRD imaging session in the same experimental conditions in the same day.

The culture medium was a normal Ringer's solution, containing 115 mM NaCl, 2.9 mM KCl, 1.8 mM CaCl2, 5 mM HEPES (4-2-hydroxyethyl-1-piperazinyl-ethanesulfonic).

2.2 Experimental and data analysis.

The experimental methods were carried out in "accordance" with the approved guidelines. The scanning micro X-ray diffraction measurements of myelin of frog's sciatic nerve were performed on the ID13 beamline of the European Synchrotron Radiation Facility, ESRF, France. A scheme of the experimental setup is shown in Figure 1A. The source of the synchrotron radiation beam is an in vacuum undulator with 18 mm period. The beam is first monochromatized by a liquid nitrogen cooled Si-111 double monochromator (DMC) and then is focused by a Kirkpatrick-Baez (KB) mirror system. This optics produces an energy X-ray beam of 12.6 KeV on a 1x1 mm² spot. The sample holder hosts the capillary-mounted nerve with the horizontal (y) and vertical (z) translation stages with 0.1 µm repeatability. The sample was scanned by using a step size of 5 µm in both y and z direction. A Fast Readout Low Noise (FReLoN) camera (1024x1024 pixels of 100x100 µm2) is placed at a distance of 565.0 mm from the sample to collect the 2-D diffraction pattern in transmission. Diffraction images were calibrated using silver behenate powder ($AgC_{22}H_{43}O_2$), which has a fundamental spacing of d(001)=58.38Å. We choose an exposure time of 300 ms for minimizing the radiation damage. The crossed bundle is of approximately 50 myelinated axons. Therefore, the diffraction frames are an average of these axons. Considering the scale of our problem, this is an acceptable average.

We measured different regions of interest (ROIs) in the central part of the nerves around their axis to minimize the capillary geometry effect on the X-ray absorption. A typical 2-D diffraction pattern with the expected arc-rings corresponding to the Bragg diffraction orders h = 2, 3, 4, 5 is shown in Figure 2A. The 2-D diffraction patterns have been azimuthally integrated to provide 1-D (Fig. 2b) angular intensity distribution I(Φ) after background subtraction and normalization with respect to the impinging beam. Information about the preferred orientation of myelin are contained in azimuthal plot, given by the integrated intensity along the around the myelin reflection from s=0.1 to s=0.3 nm-1. It shows two peaks modelled with two Gaussian whose mean value and FWHM give the average orientation of axons, $\Phi_0$ and their orientation fluctuation, $\Delta\Phi$ (see Fig. 2b).

3. Results and Discussion

3.1 Axons orientation fluctuations.

The experimental apparatus for the SµXRD is shown schematically in Figure 1a. A single nerve of frog, inserted in a capillary is placed on the sample motorized holder. The investigates nerves have a thickness of 1 ± 0.1 mm, composed of myelinated axons running vertically almost parallel to the nerve walls and perpendicular to the X ray beam. The unit of the myelin multi-lamellar ultrastructure is shown in Figure 1B. It is made of the stacking of i) one cytoplasmatic apposition (cyt), ii) one lipidic membrane (lpg), iii) one extracellular apposition (ext) and iv) another lipidic bilayer (lpg) [3-5].

The quasi-crystalline character of its ultrastructure gives rise to a diffraction pattern made of 4 centrosymmetric rings shown in the single frame picture of Figure 1a and in Figure 2a.

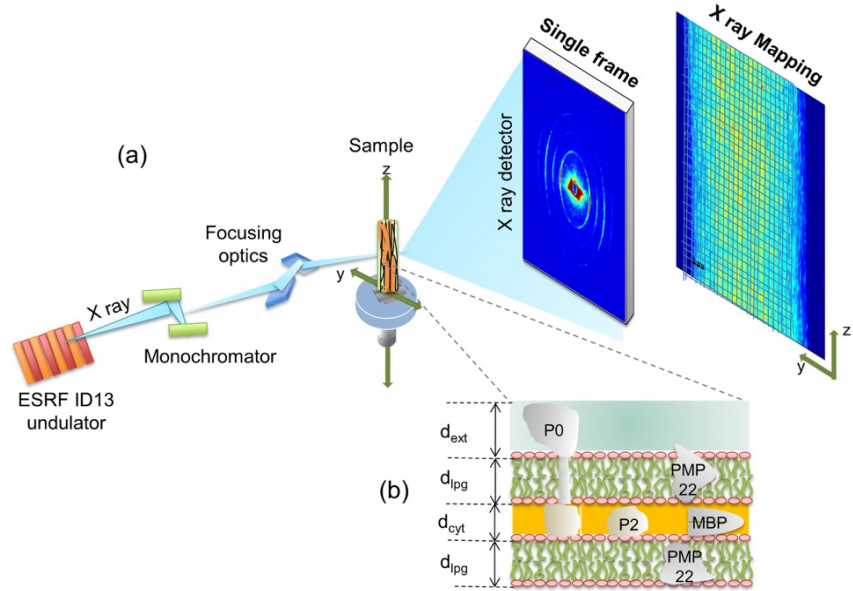

**Figure 1.** (a) Main components of the SµXRD apparatus. A typical 2D single diffraction frame, collected with the FReLon camera, shows 4 concentric arcs. Diffraction pattern measured point by point allow to build spatial maps of the extracted quantities, e.g. the total intensity. (b) Pictorial view of the protein depleted membrane layers made of polar lipid groups, lpg, with thickness $d_{lpg}$, intercalated by two hydrophilic layers: the Schwann cell cytoplasm, cyt, and the extracellular apposition, ext, with thickness $d_{cyt}$ and $d_{ext}$. respectively. The specific myelin sheath protein PMP22, P0, P2 and MBP are schematized.

The investigated samples, sciatic nerves, contain several hundreds of axons, often myelinated. The axons in the nerve could assume different orientations which changes from one illuminated spot of the sample to another. Therefore using SµXRD it is possible to map the orientation of the myelin packing.

To determine the orientation of axons we have analyzed the 1-D angular distribution of the diffracted intensity, $I(\Phi)$, as a function of azimuthal angle ($\Phi$) for each measured frame taken on each 1 µm² illuminated spot in the sample. Fig. 2 shows a typical image of the 2D diffraction pattern collected for a particular spot. Multiple curves $I(\Phi)$ collected at different spots are shown in Fig. 2b.

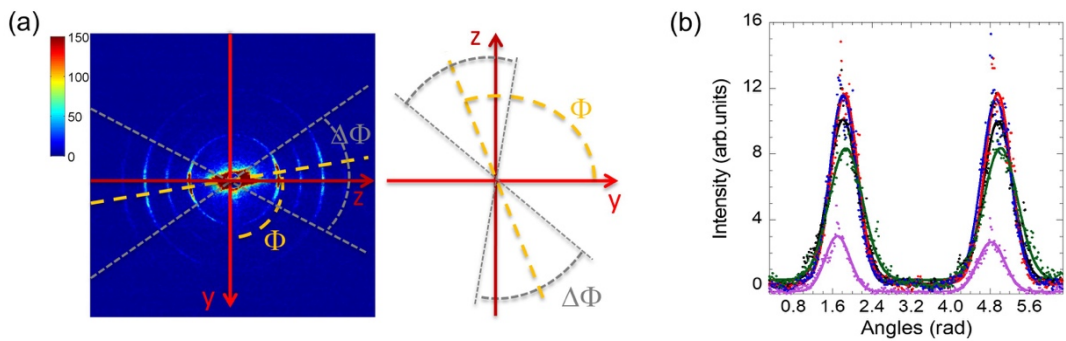

**Figure 2.** (a) Typical 2D diffraction pattern, collected with the FReLon camera, showing 4 concentric rings, with symmetrical arcs of high intensity with respect to the equatorial axis, (y). The arcs are displaced around the direction indicated by the azimuthal angle $\Phi$, counted from the y axis, and have amplitude $\Delta\Phi$. The tilting of myelin reveals an angular dispersion of the axons in the way they are arranged in the nerve. (b) $I(\Phi)$ measured at five different spots on the sample, showing the different preferred orientation of the axons, averaged throughout the tissue thickness passed by the X-ray. It shows two peaks separated by about 180° modelled by Gaussians (continuous lines). The total area under $I(\Phi)$ curve is proportional to the fraction of axons, while the FWHM represent the orientation fluctuations $\Delta\Phi$.

The arcs amplitude $\Delta\Phi$ in the 2D diffraction frames represents the spread of orientation of the lamellas, that also depends on the amount of nodes of Ranvier (also known as myelin-sheath gaps, occurring along a myelinates axon between different wrapped glia cells, where the axolemma is exposed to the extracellular space) in the illumined spots.

The curves I(Φ) show two peaks separated by 3.14 radians or 180° degree. The two peaks collected for each illuminated spot on the nerve are fitted with two Gaussians. The average FWHM ΔΦ, of the two Gaussian fits (Fig. 2b), give the orientation spread of the axons orientation in the illuminated spot. The centers of the Gaussians, $Φ_0$, give the mean orientation as described in detail in Materials and Methods. For each spot is therefore possible to collect the value of ΔΦ, normalized to mean value, $ΔΦ_0$, i.e., $ΔΦ/Φ_0$ which provides a non-dimensional parameter measuring the amplitude of the axon orientation in each spot.

3.2 Mapping of axon orientation fluctuations

The spatial distribution of azimuthal dispersion may be visualized through the spatial map of $ΔΦ/Φ_0$ in selected illuminated spots of the fresh and unfresh sample. Two typical color maps of orientation fluctuation amplitude $ΔΦ/Φ_0$ in the fresh and unfresh samples are depicted in Fig. 3a. The probability density functions, PDF, of orientation fluctuations $ΔΦ/ΔΦ_0$ for two samples are plotted in Fig. 2b. The PDF of $ΔΦ/ΔΦ_0$, for the fresh sample, shows a skewed line shape with a fat tail which we have found cam be fitted by a Levy stable distributions [22-27] shown in Fig. 3b solid red curve. Recently, the Levy stable distribution have found increasing interest in several applications in diverse fields for describing complex phenomena [22-27]. This kind of probability distribution does not have an analytical expression but are represented by a characteristic function defined by four parameters: stability index α, skewness parameter β, scale parameter γ (width of the distribution) and location parameter δ. These parameters vary in the ranges of 0<α≤2, -1≤β≤1, γ>0 and δ real. For Gaussian distribution the stability parameter becomes α=2.

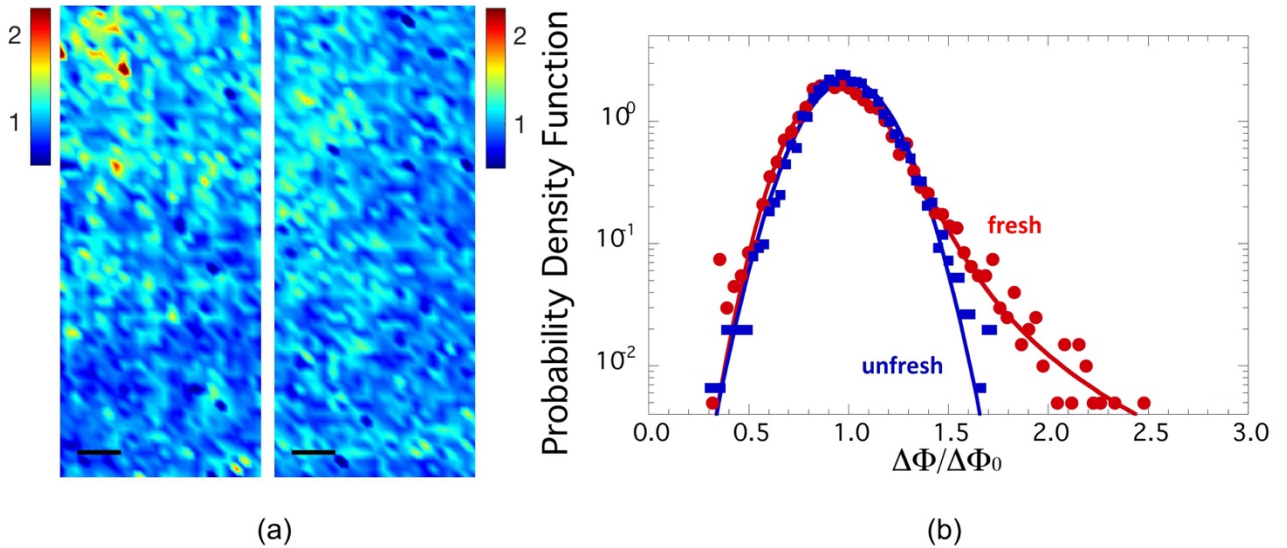

**Figure 3.** (a) Maps of axons orientation fluctuations given by ΔΦ normalized to their mean value $ΔΦ_0$, in an area of 300x125 μm$^2$ of the fresh state, to the left, and unfresh state, to the right, of axons in the sciatic nerve of a *Xenopus Laevis.* (b) Probability density function (PDF) of the orientation fluctuations $ΔΦ/ΔΦ_0$ for fresh and unfresh sample in semilog plot. We note the fat tails of the PDF in the fresh sample fitted by a Levy stable distribution.

Here we have used basic functions in the numerical evaluation of these parameters as described by Liang and Chen [28]. The Levy fitting curve, indicated by the red line in Fig. 3b, gives the following parameters of the Levy distribution:
i) stability index of 1.79 (< 2),
ii) location index 1
iii) skewness: 1
iv) scale parameter : γ of 0.1413.
Particularly interesting are the data in the PDF with higher angular distribution values. Indeed, this PDF shows a long tail, which emphasizes the presence of rare events of axons interacting between them with high fluctuations ($\Delta\Phi/\Delta\Phi_0 = 2.5$), corresponding to $\Delta\Phi$ of about 70°. These results could be indication for the system's vitality in fresh state of sciatic nerve when the nerve has been just extracted.

In order to check if these fluctuations are an intrinsic feature of the functional state of myelin, we have measured an aged state of the unfresh myelin of a sciatic nerve left 18 hours in a Ringer solution after the dissection, without oxygenation and addition of ATP. In this unfresh sample the PDF of $\Delta\Phi/\Delta\Phi_0$ follows Gaussian trend, blue line in Fig. 3b. Therefore, comparing the PDF of $\Delta\Phi/\Delta\Phi_0$ for fresh and unfresh sample, we have noted immediately that the probability distribution of the orientational fluctuations of axons, $\Delta\Phi/\Delta\Phi_0$ becomes Gaussian in the unfresh nerve, losing the fat tail of the fresh sample. Indeed, Levy distribution parameters here become:
a) stability index of 2,
b) scale parameter : $\gamma$ of 0.1315

the stability parameter of the equal 2 indicates a Gaussian distribution. Furthermore, the scale parameter decreases compared to the fresh sampl, which identifies a decreasing of interaction between axons.

The same behavior has already been observed for the conformational parameter of myelin [10]. Thus, it has been shown that this behavior is replicated both for supramolecular structures at the nanoscopic and mesoscopic levels.

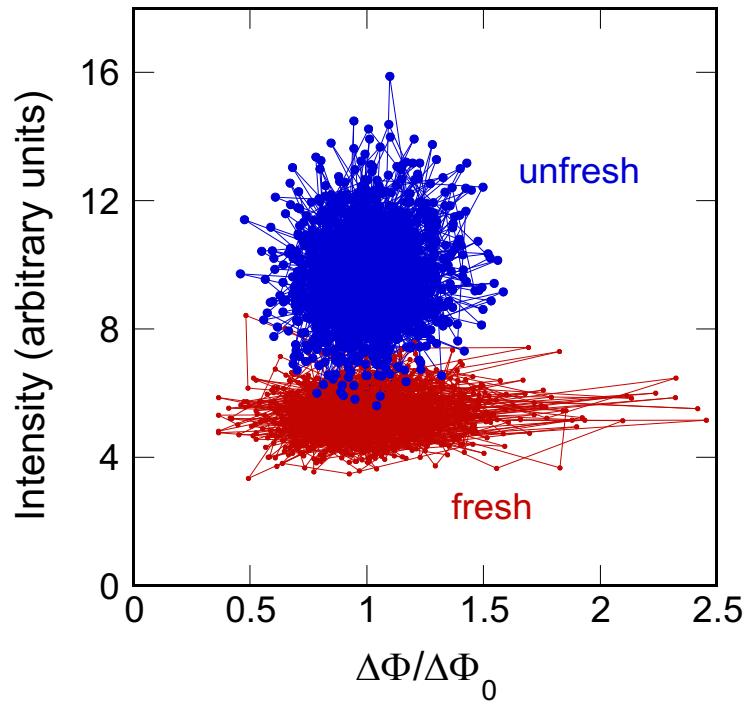

**Figure 4.** Scatter plot of diffracted intensity as a function of $\Delta\Phi/\Delta\Phi_0$. In this phase diagram the larger orientational fluctuations and lower diffracted intensity characterize the functional fresh nerve.

To get deeper insight in the geometrical fluctuations of the myelin ultrastructure we build here a phase diagram of total diffracted intensity vs. orientation fluctuations. The intensity of the XRD signal can be associated to the degree of ordering of the myelin lamellae. Going from the fresh to unfresh state we observe the increasing of the intensity in the unfresh nerve, in agreement with Poccia et al. [9]. At the same time here we observe the larger orientation fluctuations for the fresh sample. This behavior is summarized in Fig. 4, where we have plotted the Intensity vs $\Delta\Phi/\Delta\Phi_0$ for each illuminated spot, indicating the occurrence of a crystallization-like process in the aged myelin membranes, due to *a larger rigidity* in the unfresh state.

**4. Conclusion**

Understanding structure-function relationships of biological matter and new functional materials is often difficult due to their dynamic heterogeneous structure and composition at micron and submicron scale. This makes standard experimental probes probing the average structure often not adequate for the visualization of this heterogeneity, requiring high spatially resolved probes. Biological tissues are typically intrinsically heterogeneous; indeed nowadays new features and properties have been visualized using scanning methods with high spatial resolution such as Atomic Force Microscopy [29-31], Confocal Microscopy [31], Scanning Electron Microscopy [32] and Scanning micro X ray diffraction [9-10]. Correlated spatial structural fluctuations in in biological systems have been correlated with the emergence of quantum coherence in biological matter [32,33], in photosystems [34] and intrinsically disordered proteins [35,36] in as in lamellar conductors showing quantum coherence [37-40] where non-Euclidean spatial geometries for signal transmission could emerge from a correlated disorder [16,40]. Biological matter in the living state are heterogeneous systems far away from thermal equilibrium [41]. It has been proposed that living matter is made of open dissipative systems in a non-equilibrium steady state tuned close to a critical point with typical features of quantum criticality [42]. This proposal predicts that near a quantum critical point the structure of biological system should show a displays generic scale invariance and multiscale heterogeneity like critical opalescence [43].

Using scanning micro X ray diffraction measurements we have found that the fluctuations of axons in the nerve present a Levy stable distribution that is a general statistical property of complex signals deviating from normal behavior and presenting a correlated disorder. The presence of a correlation degree in the apparent disorder in biological system and its quantitative measure is of paramount importance. Here, we exploit the advances in X ray synchrotron radiation sources and optics for visualizing at micrometric scale the spatial distribution of both the orientation and the order degree fluctuations of myelin in a single sciatic nerve of a frog. The degeneration of the system due to the aging shows the loss of the correlation and a transition of the system in a random uncorrelated disordered state. This is described by the Levy to a Gaussian distribution of the orientation fluctuations accompanied by an increasing of the diffracted intensity related to the static order and thus the rigidity of the axons. The potentiality of our approach results well suited to describe the aging process of the unfresh nerve with a quantitative characterization of both spatial orientations of axons and their rigidity

In summary, we demonstrate here the feasibility of the use of SµXRD for non-invasive determination of spatial fluctuations of biomaterials which provide unique information on the correlated disorder in biological systems [18]. From measurements of the statistical axon orientation distribution we got the following main evidence: i) the fluctuation's distribution of the orientation of axons in the sciatic nerve of frog *Xenopus leavis*, in the native fresh state, is described by the Levy distribution, with large angle fluctuations. The power low fat tail behavior with rare events of spatial orientation correlation could be the experimental smoking gun of quantum critical fluctuation in the living state of myelin [41-43]. As the system start to degenerate by aging, the correlation degree of the disorder vanishes and the out-of-equilibrium living state ends up in a random uncorrelated disordered state at equilibrium at room temperature. This experiment provides a first proof of feasibility to investigate criticality in living matter opening new perspectives to be explored by further investigations on the dynamics and time evolution of spatial fluctuations of the myelin ultrastructure. The quantitative characterization by SµXRD of the correlated disorder can open many prospective and may have impact on the characterization of acute injury to peripheral nerves and its treatment.

**Acknowledgments:** The authors thank Manfred Burghammer, the ID13 beamline staff at ESRF, Grenoble, France and A. Popov of the Grenoble Institute of Neurosciences for help in the experiment.